# Understanding glassy phenomena in materials

David Sherrington[i]

Rudolf Peierls Centre for Theoretical Physics, University of Oxford
1 Keble Rd., Oxford OX1 3NP, United Kingdom
and
Santa Fe Institute, 1399 Hyde Park Road, Santa Fe, NM 87501, USA

3 October 2010

**Abstract**

A basis for understanding and modelling glassy behaviour in martensitic alloys and relaxor ferroelectrics is discussed from the perspective of spin glasses.

**I. Introduction**

There has been much activity in the last few decades in understanding, analyzing and applying knowledge of the character and properties of many-body systems with complex behaviour arising cooperatively through the combination of competitive interactions and quenched disorder, even where the individual entities, their interactions and any global constraints are simple. Example areas cover condensed matter physics, hard optimization and computer science, information science, biology and economics. They have been conceptually and technically studied and related through statistical physics, which has itself undergone major stimulation and development in the process [1].

Much progress has been made, both experimentally and theoretically, within the area of magnetic alloys exemplified by spin glasses and simple models devised to capture their essence [2]. The mathematical techniques and concepts that have been developed in spin glass theory have led to several valuable applications in the other areas outside of conventional condensed matter physics mentioned above [3], as well as in probability theory [4]. This chapter is concerned with condensed matter but in systems where the interest is in structural rather than magnetic behaviour. It uses phenomenological arguments to employ knowledge of the magnetic systems and their models to gain insight into, explain and anticipate behaviour in martensitic alloys and relaxor ferroelectrics, as well as to consider how the structurally deformable systems can provide "laboratories" to examine novel issues less accessible to real magnetic systems and suggest new problems for study in statistical physics[ii].

---

[i] Caveat: The author is not a materials scientist, but a theoretical statistical physicist concerned with modelling and understanding complex cooperative behaviour in disordered and frustrated many body systems in idealized contexts in a number of application areas. He makes no claim to expertise in the literature of the materials systems discussed in this article, but hopes that his complementary perspective can be stimulating.

**II. Spin glasses: a brief review**

Experimental spin glasses [5] are alloys of magnetic and non-magnetic ions, exhibiting frozen magnetic behaviour without periodic order, preparation–dependence, rejuvenation[iii], memory[iv] and aging [6], features generically described as "glassy". They can be metallic (*e.g.* $Au_{1-x}Fe_x$) or insulating/semiconducting (*e.g.* $Eu_xSr_{1-x}S$), $x$ giving the concentration of magnetic atoms and the spin glass features occurring for $x$ less than (system-dependent) critical values. They are commonly of substitutional character, i.e. with periodic lattice structure but random site occupation, but this is not essential. The characteristic ingredients believed to lead to their unusual cooperative behaviour are competition (or 'frustration') between different microscopic spin-interactions (some separations favouring ferromagnetic pairing, others anti-ferromagnetic pairing) and spatial (atomic) disorder, quenched on relevant timescales.

Their glassy cooperative behaviour is a consequence of the existence of many metastable macroscopic states without periodic order and with significant barriers to moving from one such state to another, with hierarchical organization and preference-relativities changing as control parameters, such as applied fields, are varied, and with the configurational entropy of the metastable states increasing as the temperature is reduced. These features appear to be ubiquitous, given the ingredients above.

Figs 1 – 3 illustrate typical properties of spin glass alloys; Fig 1 shows two phase diagrams of temperature against concentration of magnetic atoms, showing that the order which appears as the temperature is reduced from the paramagnetic state is periodic for large $x$ but spin glass for smaller $x$; Fig 2 shows results of a typical

---

[ii] The style will be tutorial/expository rather than attempting to give all historical originality credits.
[iii] Rejuvenation refers to a situation in which, after a perturbation, a system starts a process anew as though previous events had not occurred. In spin glasses it is observable in $\chi"$, which decays with time, where a sudden reduction in the temperature after decay at the higher temperature causes it to return quickly to a higher value and then start to decay again. See ref [11], also E.Vincent in ref [6].
[iv] Memory refers to a system storing knowledge of its history. For example, in the example of the previous footnote, a further sudden resumption of the earlier higher temperature makes $\chi"$ jump to the value it had just before the temperature was reduced. See ref [11], also E.Vincent in ref [6].

experiment demonstrating preparation-dependence and non-equilibration, and implying the metastability discussed above – it shows the differences in the susceptibility (magnetization/field) measured by applying a field only after cooling (ZFC, zero field cooled) and that obtained by cooling in the field (FC, field cooled); Fig 3 shows rejuvenation and memory in an experiment in which the out-of-phase susceptibility is measured as a function of time during a protocol in which the temperature is stepped down and up again.

As noted, the principal qualitative behaviour of such spin glass systems is rather universal. It is captured by the simple Hamiltonian

$$H = -\sum_{\text{magnetic}\,(ij)} J(\mathbf{R}_i - \mathbf{R}_j)\mathbf{S}_i \cdot \mathbf{S}_j \quad , \qquad (1)$$

where the $\mathbf{S}_i$ label the spins and the $\mathbf{R}_i$ their locations, $J(\mathbf{R})$ is the exchange interaction and is frustrated (competitive at different ranges), and the sum is only over the sites occupied by magnetic atoms.

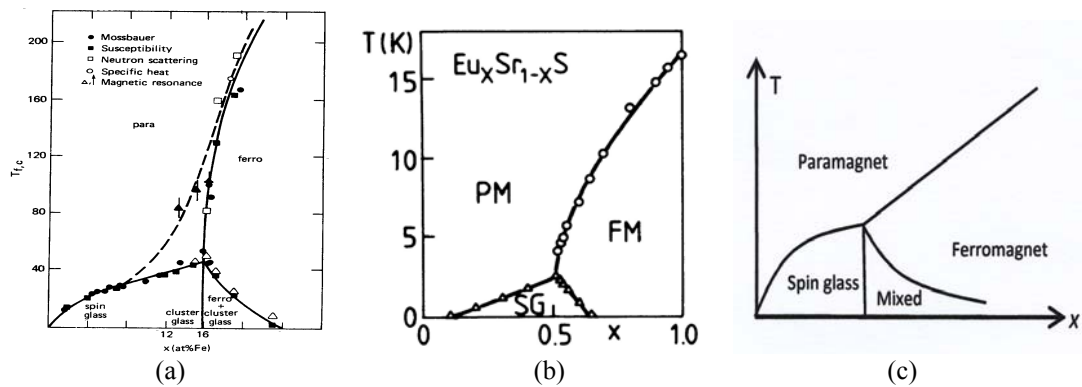

(a)            (b)            (c)

Fig.1: Phase diagrams of (a) metallic spin glass system $Au_{1-x}Fe_x$ (reprinted with permission from ref [7]; http://www.informaworld.com), (b) semiconducting spin glass system $Eu_xSr_{1-x}S$ (reprinted with permission from ref [8]; © (1979) American Physical Society; http://link.aps.org/abstract/PRL/v42/p108) and (c) mean field theory for the SK spin glass model (ref [9]) with random bonds of mean and variance both scaling as $x$.

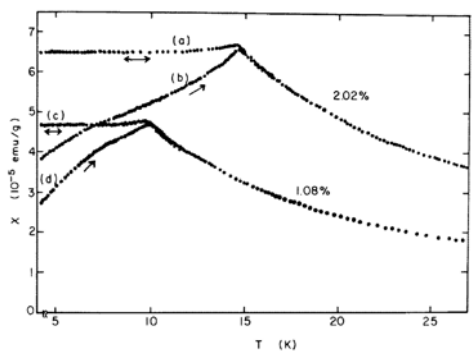 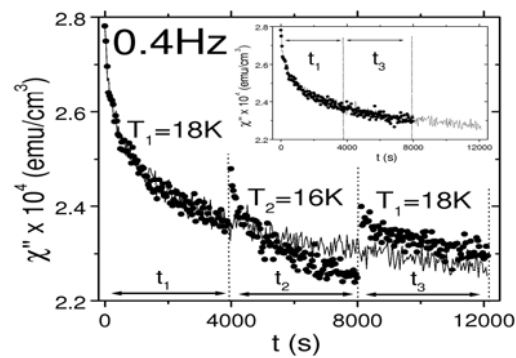

Fig.2: Susceptibilities of $Cu_{1-x}Mn_x$ as measured under field-cooling (a and c) and zero-field cooling (b and d); reprinted with permission with from ref [10], © (1979) American Physical Society; http://link.aps.org/abstract/PRB/v19/p1633

Figs. 3: Out of phase susceptibility of $Fe_{0.5}Mn_{0.5}TiO_3$ measured during field cycling as indicated, showing rejuvenation and memory; reprinted with permission from ref [11], © (2001) American Physical Society; http://link.aps.org/abstract/PRB/v64/p174204

Theoretical and computer simulational studies have played a major role in understanding spin glasses but have almost exclusively concentrated on random-bond (rather than random-site) quenched disorder, in the belief that the key ingredients are frustration and disorder and since the introduction of the model of Edwards and Anderson (EA) [12] characterized by

$$H = -\sum_{\text{all sites}} J_{ij} \mathbf{S}_i . \mathbf{S}_j \qquad (2)$$

with spins on every site but the $J_{ij}$ drawn randomly and independently from distributions $P_{sep}(J)$[v]. The EA model with only nearest neighbour interactions, uniformly distributed around $J=0$ and with Ising spins, has received much simulational study, verifying that it has features similar to those of experimental spin glasses (in the spin glass phase[vi]) and exposing a multiplicity of chaotically evolving metastable macrostates[vii], as well as many further important aspects.

The finite-range EA model is not analytically soluble but a modification to include interactions of any range with identical distance-independent probability distributions, the Sherrington-Kirkpatrick (SK) [9] model, is soluble, albeit that its solution is very subtle [3,4], and has been proven to exhibit a non-trivially evolving hierarchy of metastable macrostates, non-ergodic and aging dynamics, and the breakdown of the normal fluctuation-dissipation relation (FDR) and its replacement by a modified relation. Extensions of the SK model together with the conceptual and mathematical tools its examination has engendered have led to broad conceptual and technical application and demonstrated further subtleties, believed of relevance to understanding conventional structural glasses and several other systems.

Let us now turn to materials systems in which the interesting effects are structural rather than magnetic.

---

[v] In general the distribution depends upon the separation of the relevant sites; hence the subscript *sep*.
[vi] Note that, in accord with a common practice, we use the expression 'spin glass' to describe both a material exhibiting a spin glass phase and the phase itself.
[vii] The rejuvenation and memory effects observed in spin glasses are explainable in terms of the hierarchical yet evolving metastable state structure, with the free energy acquiring more and more nested metastability as the temperature is lowered but melting as it is raised again.

### III. Martensites

Martensitic materials [13] exhibit structural phase transitions from higher temperature phases of higher symmetry to lower temperature phases of lower symmetry, through first-order transitions. One such example, on which we shall concentrate for illustration, is from high temperature cubic austenite to a lower temperature phase of alternating planes of complementary tetragonal character, alias twins. We shall consider these systems at a phenomenological level [14].

The macroscopic behaviour of pure martensites is often considered in terms of continuum elasticity theory [13]. We also shall consider martensitic materials, including alloys[viii], as being driven by elastic considerations, but analyzed via pseudo-spin mappings and analogies employing experience from spin glasses[ix].

Our discussion will be at the level of a type of mean field/ Landau-Ginzburg free-energy theory [15, 16] and will not consider critical fluctuations. We shall mostly treat temperature simply as a means of varying effective parameters in an energy-minimization exercise. The spatial scale of the 'microscopic' variables of our modelling is coarse on the atomic scale but much smaller than macroscopic material scales.

Our starting point is to model phenomenologically the existence locally of transitions from cubic austenite to different tetragonal variants as the temperature $T$ is reduced. For further conceptual simplicity we shall initially idealise further by considering a two-dimensional analogue in which the transition is from a locally square structure to two orthogonal rectangular structures[x]. Denoting the local deviatoric strain over coarse-grained regions $i$ by $\phi_i = \varepsilon_i^{xx} - \varepsilon_i^{yy}$ this can be emulated by a local free energy

$$F_L = \sum_i \{a_i \phi_i^2 - b_i \phi_i^4 + c_i \phi_i^6\} \quad (3)$$

with the $\{b, c\}$ all positive (and of qualitatively unimportant variation) but with the $\{a_i\}$ reducing importantly as $T$ is reduced, in such a way that the local minimum changes discontinuously from $\phi_i = 0$ for large $T$ to $\phi_i = \pm \tilde{\phi}$, where $\tilde{\phi}$ is finite, as $T$

---

[viii] In fact our main interest for glassiness is in alloys.
[ix] The first recognition that there should be a spin glass analogue in martensitic alloys was by Kartha et.al. [15], looking for an explanation of 'tweed', with similarities of ideas to those discussed here, but without the direct mappings and specificity reported in this chapter, that the present author believes provide conceptual and quantitative underpinning .
[x] We shall briefly discuss extension to three dimensions later, but note at this time that since we are employing mean field theory considerations of critical dimensions caused by fluctuations are irrelevant.

is reduced through a local $T_i^{Lo}$. This can be further simplified by discretizing (and rescaling) to a description in terms of scalar pseudospin variables $S_i = 0, \pm 1$ [14, 17], with $S_i = 0$ corresponding to austenite and $S_i = \pm 1$ to the two martensitic variants, and correspondingly considering an effective local 'Hamiltonian'

$$H_L = -\sum_i D_i S_i^2 , \tag{4}$$

to be minimised. Lowering the temperature $T$ of the real system is emulated by reducing the $D$. For $D_i$ positive the local $i$-minimum is at $S_i = 0$ while for $D_i$ negative there are equivalent minima at $S_i = \pm 1$.

Next we need to include effective pseudospin interactions from one region to another,

$$H_I = -\sum_{(ij)} J(\mathbf{R}_i - \mathbf{R}_j) S_i S_j ; \quad J(\mathbf{R}) = J_{SR}(\mathbf{R}) + J_{LR}(\mathbf{R}) . \tag{5}$$

There are two types of contribution to $J(\mathbf{R})$, a short-ranged "ferromagnetic" term $J_{SR}(\mathbf{R})$ representing the inclination to follow neighbours[xi] and an effective long-range interaction $J_{LR}(\mathbf{R})$ arising through integrating out the non-ordering strains while taking account of the St Venant elasticity compatibility constraints [16, 18]; this scales as $\mathbf{R}^{-d}$ in $d$ dimensions and varies from ferromagnetic to anti-ferromagnetic depending on the angle subtended by $\mathbf{R}_i$ relative to an austenitic cell edge. In $d=2$ $J_{LR}(\mathbf{R})$ scales with distance as $R^{-2}$ with a multiplicative angular factor that yields an anti-ferromagnetic interaction at angles $\theta = (2n+1)\pi/4$ and a ferromagnetic interaction at angles $\theta = n\pi/2$ where $\theta$ is the angle subtended by $\mathbf{R}$ relative to an austenite cell edge [18].

We now consider the behaviour resulting from minimizing the total $H = H_L + H_I$, with temperature reduction emulated by reducing the $\{D\}$, and using experience of magnetic systems to make deductions about the martensites.

Let us first consider a pure system in which all the $\{D_i\}$ are the same. In this case there is a first-order transition as $D$ is lowered, from a state with all $S_i = 0$ to one with $\{S_i = \sigma_i\}$ where the $\{\sigma_i\}$ are the ground state solutions of the Ising Hamiltonian

$$H_\sigma = -\sum_{(ij)} J(\mathbf{R}_{ij}) \sigma_i \sigma_j ; \mathbf{R}_{ij} = \mathbf{R}_i - \mathbf{R}_j ; \sigma = \pm 1 . \tag{6}$$

---

[xi] This is the usual Ginzburg $(\nabla \phi)^2$ term in a spatially continuous formulation.

The transition value of $D$ is positive and given by balancing the energetic increase on taking $S = \pm 1$ in $H_L$ and the corresponding energetic reduction in $H_I$. In the lower $D$ region this ground state consists of alternating stripes of $S = +1$ and $S = -1$ at angles $\pi/4$ or $3\pi/4$ [xii]. These are the twins of pure martensite.

Now let us turn to alloys, emulated by a random distribution of the $\{D_i\}$ over the lattice. Again the criterion of whether any $S_i$ is 0 or $\pm 1$ is given by the balance of $H_L$ and $H_I$, determined self-consistently across the whole system.

For conceptual orientation it is useful to consider first a scenario where the $D$ are distributed randomly and independently across the $i$, at each site taking one of two values; with probability $(1-x)$ a large $D_0$, such that sites with this $D$ always have $S_i = 0$, and with probability $x$ a smaller $D_1$ that can be varied across a phase transition (emulating reduction in temperature). For large enough $D_1$ the ground state is austenitic (all $\{S_i = 0\}$). Lowering $D_1$ further, a transition will occur into a phase with $\{S_i = \sigma_i\}$ on the sites having $D_i = D_1$ when there is first a solution of

$$\sum_i c_i D_1 - \sum_{(ij)} c_i c_j J(\mathbf{R}_{ij}) \sigma_i \sigma_j = 0 \; ; \sigma = \pm 1, \qquad (7)$$

where $c_i = 1, 0$ for $D_i = D_1, D_0$. By comparison with eqn. (1) the second term of eqn. (7) is recognised as that of a site-disordered Ising spin glass system with the magnetic sites corresponding those with $D_i = D_1$ and with exchange $J(\mathbf{R})$. The nature of the ordered phase depends on the ground state energy of $H_{eff} = -\sum_{ij} c_i c_j J(\mathbf{R}_{ij}) \sigma_i \sigma_j$ and can be either ferromagnetic or pseudospin glass.

Currently we have no precise calculations for the ground state energies of $H_{eff}$ with the specific interaction of eqn. (2)[xiii]. However, the $(T,x)$ phase diagrams of conventional spin glasses give an indication of what to expect, since the transition temperatures at magnetic concentrations $x$ provide estimates of the corresponding ground state energies. In particular, there is a critical $x_c$ (depending on system details)

---

[xii] There has been much interest recently in stripe ordering in systems with a combination of short-range ferromagnetic and long-range power-law-decreasing anti-ferromagnetic interactions and it has been proven that the preferred order is of stripes for $d < p \leq d+1$ where $d$ is the spatial dimensionality and $(-p)$ is the power of the long range decay [19]. Stripe widths are determined by the relative strengths of the two types of interaction. Here $p = d = 2$ and the system is marginal with relevant boundary size $L$ and it has been shown that the average twin stripe width depends on $L$ (as the square root); see ref [20].

[xiii] Indeed, even with specified interactions, its evaluation is surely NP-hard [21].

separating a high $x$ periodic order regime from a lower $x$ spin glass phase, with the ordering temperature (and correspondingly the ground state energy) growing with $x$ in both spin glass and periodic phases, with a discontinuous increase in $dT_c / dx$ (and correspondingly the binding energy per non-zero spin) at $x_c$. Consequently, for the martensitic alloys we expect transitions as $D$ is reduced (i) from austenite to twinned martensite (the analogue of the ferromagnet in the magnetic examples shown) for $x < x_c$, with $x_c$ dependent on details of $H_1$, (ii) from austenite to a pseudo-spin glass frozen amorphous martensitic state for $x > x_c$, with the critical $D$ for the transition increasing monotonically with $x$ and with a positive discontinuity in $dD_c / dx$ at $x_c$. The pseudo-spin glass state would be expected to exhibit non-ergodic behaviour analogous to that found in spin glasses, such as differences between FC and ZFC uniaxial compressibilities. This behaviour was observed recently [22, 23] and the pseudospin glass state named 'strain glass'.

In reality one might expect a more quasi-continuous range of local $D$-values, particularly allowing for the coarse-graining implicit in our effective site-description. Hence it is reasonable to consider the case in which the $D_i$ are chosen independently from a distribution $P(D)$ of mean $D_0$ and standard deviation $\Delta$ and study the behaviour as $D_0$ is reduced, emulating reduction in temperature of the real materials. This will lead to different local penalties for $S \neq 0$ at different sites and hence different amounts of bootstrapped interaction energy needed to convert locally to favourable $S = \pm 1$. Specifically, any site $i$ will convert from $S_i = 0$ to $S_i = \pm 1$ at a critical $D_i$ given by $D_i = -\delta H_I^i(c)$ where $\delta H_I^i(c) < 0$ is the resultant change in value of $H_I$ with already a fraction $c$ of sites converted. The actual sign choice of $S_i$ will depend upon the specific instance of the $\{D\}$ and the states of the other $\{S_j\}$ but the magnitude of $\delta H_I^i(c)$ is expected to be dominantly self-averaging and again it can be estimated from the $(T,c)$ phase diagram of the corresponding spin glass system[xiv] or, in its absence, qualitatively from those of known spin glasses. Thus we expect the transition from austenite as $D_0$ is reduced to be to martensitic twins for $\Delta < \Delta_c$ and to pseudospin (strain) glass for $\Delta > \Delta_c$. Again, this is in accord with observation, noting

---

[xiv] i.e. random-site Ising with the same $J(R)$.

that $\Delta$ is expected to be a monotonically increasing function of the defect concentration in alloys, for example in $Ti_{50-y}Ni_{50+y}$[xv] [22], for small $y$.

Within the lower temperature region there will be transitions from twinned to strain glass as the disorder concentration is varied. Again spin glasses can be used to guide expectations. Within the (soluble) SK model this transition is at a constant $x = x_{c1}$ for all $T$ but within the ferromagnetic region there are two sub-regions; for $x > x_{c2}(T)$, with $x_{c2}(T)$ increasing from $x_{c1}$ as $T$ is reduced, the ferromagnetic phase is ergodic, but for $x_{c1} < x < x_{c2}(T)$ the ferromagnetism is non-ergodic or "mixed" (ferromagnetic-spin glass).

Turning to the martensitic systems with a quasi-continuous distribution of $D$, it can be noted that as $D_0$ is reduced more and more sites will pass the threshold of eqn. (7) and hence become magnetic sites in the effective Ising spin glass Hamiltonian. Correspondingly, the effective concentration of magnetic sites will increase as the temperature of the martensitic system decreases and the boundaries between ordered and strain glass regions will move further and further into the twinned region, yielding re-entrance, so that for $\Delta$ just greater than $\Delta_c$, where $\Delta_c$ is the critical disorder at which the transition from austenite to lower symmetry occurs, one can anticipate a sequence of phases on lowering the temperature of austenite $\rightarrow$ strain glass $\rightarrow$ mixed twins/strain glass phase $\rightarrow$ ergodic twinned martensite[xvi]. Some features of an intermediate phase and re-entrance have been seen in experiments although it is probable that the mixed phase is not truly equilibrium but rather only manifest on finite time scales. Figs 4 to 6 show the prediction and some experimental observations[xvii].

For conceptual simplicity the description above has been in terms of two-dimensional modelling. It can however be extended simply to three dimensions, for example by employing a lattice gas description $n_i = 0,1$ to indicate whether a site is austenitic or martensitic, denoting the three orthogonal tetragonal variants by Potts "spins" $p_i = 1,2,3$, and writing the pseudo-spin Hamiltonian as

---

[xv] Note that $y$ measures the density of defects compared with the pure case $Ti_{50}Ni_{50}$ whereas $x$ earlier was the density of normal (host) atoms.
[xvi] Such re-entrance has been called "inverse freezing" and has been receiving much attention in other contexts; e.g. [25].
[xvii] Note: The transitions shown between pure twinned and mixed phase and between twinned and strain glass are qualitative but both are swung towards the twinned state compared with their SK counterparts.

$$H = \sum_i D_i(1-n_i) - \sum_{(ij)} n_i n_j J(\mathbf{R}_{ij})(\delta_{p_i p_j} - 1/3). \quad (7)$$

Again $J(\mathbf{R})$ will have a short-range ferromagnetic part and a long-range part (but now, in three dimensions, going as $R^{-3}$), again with angular variation from negative to positive and favouring the usual twin planes. The consequences will be similar to those discussed above[xviii].

In fact, the original motivation for investigating a possible spin glass analogue in these systems [15] was to explain the possible origin of a different, apparently disordered state of martensitic materials that was observed as a pre-cursor above the transition to the twin structure, exhibiting a mixture of regions of different tetragonal- and austenite-distortions and known as "tweed"[xix]. The authors recognised this behaviour as analogous to the amorphous appearance of spin glasses and speculated

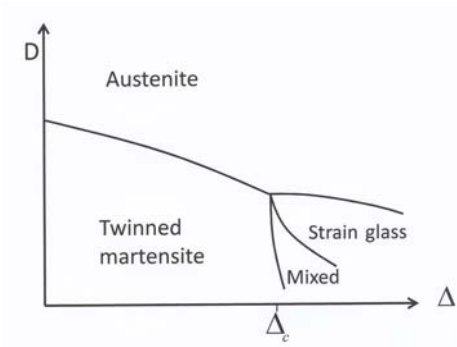
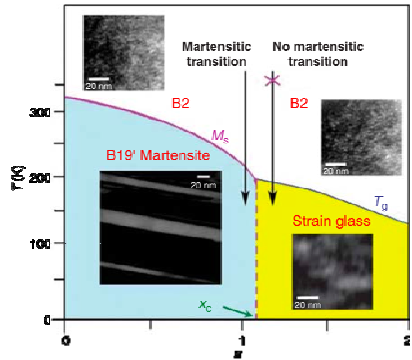

Fig 4. Qualitative predicted phase diagram    Fig 5. Phase diagram of $Ti_{50-x}Ni_{50+x}$ reprinted with permission of MRS Bulletin from ref [23], Fig 2.

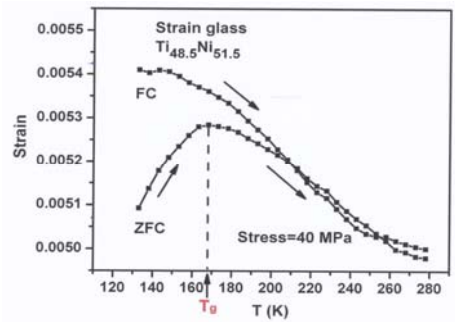
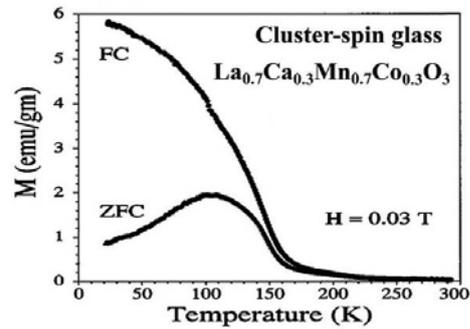

(a)    (b)

Fig 6: (a) Compressibility of strain glass $Ti_{48.5}Ni_{51.5}$ (reprinted with permission from ref.[22]; © (2007) American Physical Society; http://link.aps.org/abstract/PRB/v76/p132201), with, for comparison, (b) susceptibility of a cluster spin glass (reprinted with permission from ref. [26]; © (2007) American Physical Society; http://link.aps.org/abstract/PRB/v56/p1345).

---

[xviii] Within mean field theory, Potts spin glasses of Potts dimension greater than 2 show additional re-entrance from spin glass to ferromagnet as the temperature is reduced [27].
[xix] This was also the initial motivation that led to ref [14].

that it might be an analogous glass, recognising also that quenched randomness in the constitutive make-up was a necessary ingredient and assuming a model based on the SK spin glass with effective temperature-dependent bond-randomness and an analogue of a spin glass phase between two ordered phases emulating austenite and twinned martensite[xx]. In fact, it is now believed that the original tweed phase that is observed as precursor to the twinned martensitic phase is ergodic [28]. What was being anticipated theoretically was strain glass, as the above discussion demonstrates. It seems probable that tweed is actually a non-equilibrium precursor. Bounds for its existence follow qualitatively from the spinodals of the Landau-Ginzburg free energy obtained from the combination of the soft-spin local $F_L$ and a soft-spin extension of $H_I$ (with the discrete-valued $S$ replaced by soft $\phi$). It seems probable that a metastable tweed precursor will also occur at temperatures above the strain glass in its region of defect space but it remains to consider it further theoretically.

One of the characteristic features of martensitic materials is one-way shape-memory in which a shape that is imposed in the high temperature austenitic phase, e.g. by plastic distortion or moulding at the time of preparation, is easily removed (or further distorted) by the application of only weak force in the twinned phase beneath the austenite-martensite transition, but reappears on heating back above that temperature. The usual pictorial description is in terms of (i) any high-temperature imposed macroscopic shape being maintained under cooling through the martensitic temperature while simultaneously the structure distorts microscopically into an equal mixture of twin types, with (ii) further distortion in the twinned phase easily achievable by redistribution of weights among the twin types and (iii) returning to austenite and the original shape on heating. One might wonder about the relationship of this effect to the memory effects observed in spin glasses. In fact, however, it is readily explainable without glassiness but does require going beyond hard pseudospins to deal with the plastic distortions in the martensite phase. The fact that martensitic twins are quite soft to distorting stresses which leave remanent strains demonstrates that the minima in the Landau-Ginzburg free energy are shallow, as also does the superelasticity exhibited above the martensitic phase transition temperature. Nevertheless, it would be interesting to look for analogues of spin glass non-

---

[xx] Ref. [14] also assumed that the origin of tweed was quenched disorder but, as above, locally in a system with frustrated but not necessarily disordered exchange. In fact, the consequence is strain glass.

ergodicity, such rejuvenation and memory effects of Fig. 3, which require a hierarchy of chaotically evolving metastable states. Indeed two-way shape memory was demonstrated in a computer simulation [29] and argued to be such a spin-glass like manifestation.

It should be emphasised that temperature has only been introduced implicitly through the variation of parameters in the Ginzburg-Landau free energy, particularly through the variation of the mean of the distribution of effective $\{D\}$. This would need to be noted if one wished to find the solutions of the model above by computer simulations; in particular, if one wished to investigate the ground state energy of $H_L + H_I$ by simulated annealing then one would need to employ another, artificial, annealing temperature $T_A$ and reduce it to zero. A study of real thermal fluctuations would require modelling in terms of a real Hamiltonian, as opposed to this Ginzburg-Landau phenomenological emulation.

**IV. Relaxors**

Another set of materials that undergo interesting structural deformation with apparently glassy-like non-ergodicity are the so-called ferroelectric relaxors [30, 31, 32]. Here we shall concentrate on systems epitomized by $PbMg_{1/3}Nb_{2/3}O_3$ (usually abbreviated as PMN)[xxi] and its alloys with $PbTiO_3$ (abbreviated as PT). They exhibit

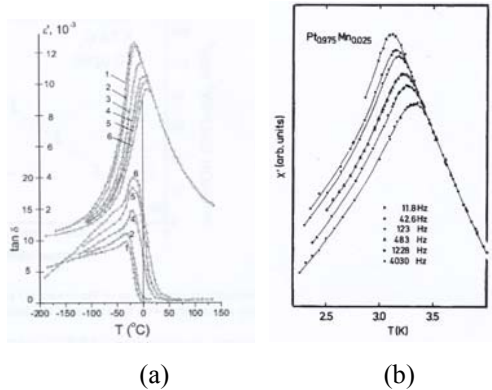
(a) (b)

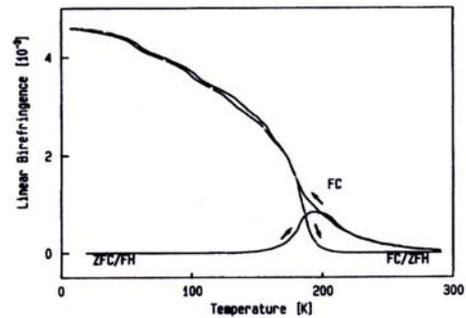

Fig 7: (a) Upper part: Temperature-dependence of real part of the dielectric permittivity of PMN for a range of frequencies, increasing through curves curves 1-6 (reprinted with permission from ref [34]; © (1961) American Institute of Physics);
(b) Frequency-dependence of real part of magnetic susceptibility in $Pt_{0.975}Mn_{0.025}$ (reprinted with permission from ref [35]).

Fig 8: Linear birefringence of PMN in (001) plane induced by an electric field of 3kV/cm along [110], under conditions of field cooling (FC) and zero-field cooling/field heating (ZFC/FC) (reprinted with permission from ref [33]; © (1992) American Physical Society; http://link.aps.org/abstract/PRL/v68/p847)

---

[xxi] Another example is $PbZn_{1/3}Nb_{2/3}O_3$ (abbreviated to PZN).

several features similar to spin glasses, including non-ergodicity as the temperature is lowered beneath a transition temperature [33]. Fig 7 gives a comparison of the temperature- and frequency-dependence of the dielectric permittivity of relaxor PMN[xxii] and the magnetic susceptibility of spin glass $Pt_{0.975}Mn_{0.025}$, while Fig 8 shows FC and ZFC measures for PMN to be compared with those of a simple spin glass shown in Fig 2. Both of these comparisons suggest similarities and hence the existence of a multiplicity of metastable macrostates in the relaxors[xxiii].

PT is a member of a class of perovskite ionic crystals (see Fig 9)

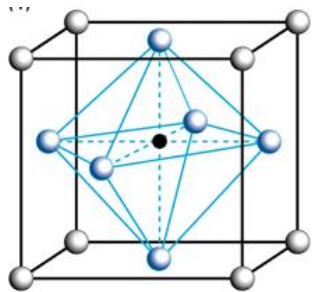

Fig 9. Perovskite structure $ABO_3$. Here A sites are at the corners of the cube, a B site is at the centre of the cube and the O sites are at the centre of the faces.

characterised by the structure $ABO_3$ in which the A have charge +2, the B have charge +4 and the O have charge -2; typical examples of A are Sr, Ba and Pb; of B, Ti and Ta. Their more detailed structures are determined by the balance of their forces (short-ranged forces related to the sizes of the ions and longer-range Coulomb forces of both signs) and at lower temperatures they exhibit ferroelectric or anti-ferroelectric distortions to lower symmetry, depending on the specific system. In the case of PT the low temperature order is ferroelectric. PMN however is a substitutional alloy with the Ti ions (of charge +4) replaced by a mixture of 1/3 Mg (of charge +2) and 2/3 Nb (of charge +5), believed distributed quasi-randomly but maintaining coarse charge neutrality. $PT_{1-x}PMN_x$ alloys[xxiv] span the range from fully periodic to maximally random.

Let us now turn to modelling. Our philosophy will be to take a bare Hamiltonian $H_0$ to characterize minimally the displacement properties of pure PT,

---

[xxii] Relaxors are so-called because of this significant frequency-dependent permittivity peak behaviour.
[xxiii] The feature of frequency-dependence of the peak in the real part of the dielectric permittivity or the magnetic susceptibility as a function of temperature decreasing with decreasing frequency is interpretable as reflecting the range of characteristic barrier penetration times, with higher barriers taking longer to surmount.
[xxiv] The usual convention in the field is to write PMN-PT but we shall use PT-PMN here in keeping with our perspective of disordering a pure matrix.

with a perturbation Hamiltonian $H_1$, again minimal, characterizing the perturbations caused by alloying with $Mg_{1/3}Nb_{2/3}$ in place of Ti.

In general $ABO_3$ systems, as the temperature is lowered, there can be displacement from the high temperature pure perovskite structure of any of the constituent ions. There are competing forces at play with their relative strengths determining the actual low temperature states, due to the mixture of signs of charges present as well as the normal short-range atomic forces, and this is evidenced by the fact that some $ABO_3$ are ferroelectric, some anti-ferroelectric. However (i) our principal interest is in the effects of substitutionally disordering the pure material by replacing B ions with a mixture of ions of different charges, (ii) in PT and PMN the main displacement is observed to be of the Pb ions, and (iii) it is known that PT is ferroelectric in a <111> direction. Hence, as a first approximation, we shall take the displacement structure of the pure low temperature PT to be modelled simply in terms of Pb displacements via an idealized Hamiltonian

$$H_0 = \sum_i (rS_i^2 + uS_i^4) - \sum_{(ij)} J(\mathbf{R}_{ij}; \mathbf{S}_i, \mathbf{S}_j) \qquad (8)$$

where the $i$ label Pb ions, the $\{\mathbf{S}_i\}$ are their displacement vectors, allowed to vary continuously in length and direction, $r$ is positive reflecting the elastic energy to move Pb atoms from their normal (high temperature/high symmetry) positions, $u$ is positive[xxv], limiting displacement, and the $J(\mathbf{R};\mathbf{S},\mathbf{S})$ term is an anisotropic exchange term favouring ferroelectric ordering along <111> directions and large enough to overcome the positive $r$ and drive cooperative ferroelectric ordering when the temperature is reduced. In an obvious magnetic analogy we shall refer to the $\{\mathbf{S}_i\}$ as (soft) spins.

Within this picture, the perturbation to PT caused by substitutionally alloying $Mg_{1/3}Nb_{2/3}$ in lieu of Ti can be considered by adding to $H_0$ an extra Hamiltonian contribution corresponding to extra charges -2 at locations occupied by Mg ions and charges +1 at locations occupied by Nb ions, with long-range consequences; together with short-range perturbations due to the different ionic radii of Ti, Mg and Nb. Let us concentrate on the effects of the charge perturbations. Since the Pb are charged, the perturbing extra charges on B sites will lead to additional Coulomb forces trying to

---

[xxv] Note that this is in contrast to the case of the martensitic materials discussed above, and implying a continuous transition, although one could easily modify to a negative $u$ and include a positive 6th order term to bound the Hamiltonian if one wished to allow the possibility of a first order transition.

displace the Pb ions from their PT positions. Ignoring any displacements of the Mg or Nb themselves from the equilibrium Ti positions, these charges lead to a perturbing Hamiltonian $H_1 = \sum_{i\alpha} \mathbf{h}_{i\alpha} . \mathbf{S}_{i\alpha}$ where the $\mathbf{h}_{i\alpha}$ are effective 'random fields' experienced at A-sites $i$ due to the extra charges at B-sites α, taking the form $\mathbf{h}_{i\alpha} = -2\mathbf{g}_{i\alpha}$ if there is an Mg ion at α and $\mathbf{h}_{i\alpha} = +\mathbf{g}_{i\alpha}$ if there is a Nb ion at α, where $\mathbf{g}_{i\alpha}$ is a vector that points in the direction from $i$ to α and whose magnitude scales with separation as $(R_{i\alpha})^{-2}$; there is no contribution ($\mathbf{h}_{i\alpha} = 0$) from α-sites where the Ti ions are not substituted.

There has been a lot of interest in random field problems, paralleling the interest in random exchange problems typified by EA-like spin glasses [36][xxvi]. For simple ferromagnets with continuous vector spins it has been shown that uncorrelated random fields destroy the long-range order at dimensions less than four [37] due to the formation of domains. Much of the theoretical interest has however been in the random-field Ising model (RFIM), with uniaxially restricted spins, non-negative exchange interactions and uncorrelated random fields, for which the critical dimension for domain formation is 2. For this system it was predicted [38] that in 3 dimensions and at intermediate temperatures there would be non-ergodic behaviour analogous to that of spin glasses, although a clear solution is still elusive and this non-ergodic state has been shown recently not to be the true equilibrium solution for a system with only ferromagnetic (or zero) interactions [39].

But note that here (i) the fields are not certainly not uncorrelated, both because of the need to maintain approximate local charge neutrality and also because of the high correlation of the field directions experienced by pairs of Pb on either side of a Mg (both towards the Mg) or on either side of a Nb ion (both away from the Nb) and (ii) while the "bare" system is ferromagnetic, the full interaction term in the effective spin Hamiltonian (8) also has anti-ferromagnetic elements as a function of {**R**} that could become relevant under inhomogeneous perturbation (as is the case for random site spin glasses and martensitic systems). Furthermore, we know from conventional glasses that non-equilibrium glassy states can occur easily in practice even when the minimum energy state is crystalline.

---

[xxvi] Recall that most experimental spin glasses have site disorder but frustration in their exchange.

Experimental implementation of uncorrelated random-signed fields in conventional magnetic systems has not proven possible, so experimental studies have studied instead random Ising anti-ferromagnets in the presence of uniform fields, utilising a mapping to corresponding ferromagnets in random fields (directed oppositely on the sites of the two 'hidden' anti-ferromagnetic sub-lattices). These have exhibited effects of non-ergodicity [40].

Relaxor ferroelectrics, such as PT-PMN, however, would seem to provide naturally effective random fields at random sites on the Ti lattice, subject to mesoscopic charge neutrality but not related to further implicit sublattices, on top of a bare non-disordered effective Hamiltonian that leads to ferromagnetic ordering when all its sites are occupied, but with anti-ferromagnetic interactions too. They therefore have the potential to be very interesting laboratories to study fundamentals of random-field problems. They do have vector 'spins' but also with anisotropy that prefers the <111> orientations and can effectively change the spin character from continuous to quasi-discrete as the temperature is lowered. That quasi-discreteness is not however simple Ising (with two states) but rather has eight possible equivalent orientations, as also do the directions for the strongest (nearest-neighbour-effected) random fields[xxvii].

Experimentally, these relaxors have received extensive study with many interesting observations and deductions but without a consensus of understanding or detailed theory. Pure PT exhibits a ferroelectric phase transition at around 700K while PMN exhibits more than one 'characteristic' temperature, a so-called Burns temperature at around 620 K marking an onset of deviations from a simple extrapolation of higher temperature properties, and two 'transitions' [32], one around 420K but maintaining ergodicity and another around 220 K [33] heralding the onset of no-ergodic behaviour. It has been suggested that these 'transitions' indicate respectively, first, a random-field transition but with the spins still having enough vector freedom to distort continuously transversely and, second, behaviour analogous to that of RFIM as a consequence of freezing out of the transverse continuous freedom through the anisotropy favouring the <111> directions and hence the onset of spin angular discreteness [32]. Experiment indicates that (small) domains grow as the temperature is lowered in the higher of these regimes (and probably already starting

---

[xxvii] Effective random fields from further B atoms have intermediate orientations.

from around 600K), as might be anticipated from Imry-Ma theory for the regime where anisotropy is less effective in hindering angular deviation.

Some authors have suggested that these systems should be considered as random field. On the other hand, several authors have proposed that the non-ergodic phase of PMN and PMN-PT should be envisaged as a kind of spin glass freezing of the nano-domains through effective random domain exchange. In view of the large strength of the perturbing random fields (arising from their Coulomb origin) it is indeed likely that their effects will often be greater than that of some of the effective exchange terms and require account to be taken of longer range anti-ferromagnetic aspects of $\sum_{(ij)} J(\mathbf{R}_{ij}; \mathbf{S}_i, \mathbf{S}_j)$ [xxviii]. We have already noted in both the examples of spin glasses and martensitic alloys that the addition of local randomness to a pure but frustrated 'bare' Hamiltonian can lead to behaviour analogous to a random bond system and there are several other possible examples. Furthermore, there are often different ways to write Hamiltonians for different apparent emphasis and different ways to choose how to separate bare and perturbation parts of a full (even minimal) Hamiltonian. For example, the idealization of the PT-PMN system discussed above can be written to emphasise random spin correlation in the perturbation (restricted for illustration to nearest neighbour) as

$$H_0 = \sum_i (rS_i^2 + uS_i^4) - \sum_{(ij)} J(\mathbf{R}_{ij}; \mathbf{S}_i, \mathbf{S}_j) - \sum_{(kl)} c_{kl} a_{kl} (\mathbf{S}_k - \mathbf{S}_l).(\mathbf{R}_k - \mathbf{R}_l) \qquad (9)$$

where (kl) are pairs of near neighbour Pb sites, $c_{kl} = 1$ if the Ti between k and l is replaced, otherwise $c_{kl} = 0$, and the $a_{kl}$ have opposite signs (and different magnitudes) for replacement by Mg and Nb ions.

There are also several other examples of modified $ABO_3$ in which the randomizing or alloying is different and can be expected to lead to different pseudospin pictures; for example a B can be replaced by a B′ with the same charge so there is no Coulomb contribution to the perturbation $H_1$, as in $PbZr_xTi_{1-x}O_3$, or even more perturbed by replacing some A by ions of higher charge together with free electrons, such as replacing some Pb by La. Thus even the minimal model may need to be different in different cases. However, despite these different origins and details one is led to suspect that understanding of any non-ergodicity in these materials is to be found in terms of the multiple metastable state picture.

---

[xxviii] Also, in this connection it is probably appropriate to draw attention to recent studies of dipolar glasses with local anisotropic preferred-axis disorder [41].

Finally, in this section, we might note that there are other types of relaxor ferroelectrics. For example $Sr_{-0.61-x}Ce_xBa_{0.39}O_6$ has been proposed as a uniaxial relaxor realizing the RFIM in a materials analogue [42]. We shall not, however, pursue these further here.

**V. Models, simulations and analysis**

Minimalist models, clever computer simulation and often-subtle analytical studies have played important roles in helping understand spin glasses and generalise concepts. Real experimental systems have many controlling parameters and many variables, of different degrees of importance for determining behaviour, making complete modelling and simulation potentially confusing and difficult to perform on large enough systems. Hence the desire to simplify as much as possible while maintaining what are believed to be the most important ingredients, revising such beliefs in the light of comparison of predictions and observations. Simulations on such pared-down models enables larger systems to be studied under conditions that are known and also permits potentially instructive measurements for which no real experiments have yet been devised – an example of such a measurement that proved of great value in studies of spin glasses is of cross-correlations of two systems evolving simultaneously with the same control rules but different instances of stochastic noise.

Analytic solutions are essentially impossible for large disordered and frustrated systems, except for certain typical properties of systems with only infinite-range interactions drawn independently and identically from the same distribution, epitomised by the SK model and its extensions, including local randomness also drawn independently from site-independent distributions. These latter have, however, been very instructive in forming conceptual pictures and devising new procedures,

The model of eqn. (1) is a simple emulation of a real experimental system, but in fact such site-disordered systems have received little simulation or analysis; these have instead concentrated on models with bond-disorder, such as that of eqn (2), because they are argued to have the same qualitative character within the spin glass phase. One could take the same perspective in discussing the martensites above if one is principally interested in the properties of the strain glass phase. Thus one might employ the model Hamiltonian

$$H = \sum_i D_i S_i^2 + \sum_{(ij)} J_{ij} S_i S_j; S = 0, \pm 1 \qquad (10)$$

where the *D* and the *J* are drawn from simple distributions (such as Gaussian or top hat functions). This model has been studied analytically for the (soluble) infinite-ranged case [43, 44, 45], exhibiting the anticipated amorphousness of the strain glass state. A model combining this with a term of the form of eqn. (6) to emulate the transition from twinned martensite to strain glass has also been simulated [45]. These models, however, effectively put in ingredients guaranteed to yield strain glass. It would be interesting to minimize numerically $H = H_L + H_I$ of eqns. (4) and (5)[xxix] and show the emergence of the strain glass, but it is difficult to anticipate any result other than that phenomenologically deduced above. Indeed, simulations of the initial elasticity model have clearly demonstrated features analogous to tweed, as well as showing martensitic stripes [15, 18, 29, 47].

These considerations do suggest several other models and experimental systems as being of potential interest[xxx]. For example, there has been much interest in the cooperative formation of striped phases in many pure systems [48]. The above arguments suggest that dilution of such systems might often lead to transitions to 'amorphous' glassy order. This could be the case with various 'spin' types and one could consider examples with second-order or first-order transitions, driven from their high symmetry phases by changes of anisotropy, thermal or quantum fluctuations.

The situation for the random field model suggested for PT-PMN is more difficult to consider analytically. For example replacement of the finite-range RFIM problem by one with uniform infinite-range exchange leads to triviality and thus mitigates against a useful simple mean field solution. Nor has the RFIM received much simulation since this is made difficult by the discreteness of the spins. There is consequently much more uncertainty about glassiness in random field problems. The observation of non-ergodicity in PMN and PT-PMN suggests, however, that it could be useful to consider minimal model simulation for these systems as random field ones but appropriately correlated as discussed above. Details of the correct exchange and its anisotropy in eqn. (8) are not immediately obvious and their realistic inclusion would also complicate simulation, but it could be interesting to simulate a model with a simplified form, perhaps first with a purely ferromagnetic exchange and local

---

[xxix] For example, by simulated annealing or extremal optimization [46].
[xxx] Indeed some will have been studied but no attempt at a survey is made here.

anisotropy, moving to a competitive form with also an anti-ferromagnetic exchange at greater separation as necessary, looking for an onset of non-ergodicity as the temperature is reduced and anisotropy frozen out. But this is beyond the scope of the present article.

There has been an attempt to emulate the relaxor system with an infinite-range random bond and random field model [49] but again this effectively pre-empts the conclusion, while the even-qualitative validity remains undemonstrated.

## VII. Conclusion

In this article it has been argued that experience in spin glasses and random magnets can provide useful perspectives for considering the origins of complex glassy behaviour in materials, although no claim is made of completeness of picture or of full originality. In keeping with the tradition of statistical physics the approach has been to try to simplify the materials problems to provide minimal models for basic understanding, to consider the implications of such mapping through comparison with known spin glass systems, to be followed eventually (not here) by extension to greater reality. The glassiness has consequently been identified in terms of the complex metastable state structure of spin glasses and high-disorder random field systems.

Specifically it has been argued that martensitic alloys with compositional defects behave like spin glass systems with first-order phase transitions as the temperature is lowered from the high temperature high symmetry austenite phase to a lower temperature periodic (twinned) phase for lower levels of defect concentration, or to a non-ergodic spin-glass like phase for higher levels of defect concentration. Furthermore it has been suggested, from a mapping to a picture of bootstrapped effective Ising[xxxi] spins, that if the quenched disorder has a quasi-continuous anisotropy strength distribution then the transition from twinned to strain glass phase as a function of temperature and defect concentration should be re-entrant. This feature implies that martensitic alloys could provide a useful "laboratory" for readily studying effective concentration variation across a periodic-spin glass transition as a function of temperature of the martensitic material rather than requiring the making new alloys of different composition.

---

[xxxi] or Potts

It has also been argued that PT-PMN relaxor ferroelectrics are most minimally modelled as random field problems and might provide a useful experimental laboratory to study such model systems without needing to employ gauge mappings from random anti-ferromagnets in uniform fields, with their necessarily uniaxial fields of uniform strength and with the quasi-randomness tied to underlying sublattice structure. On the other hand it has also been pointed out that these systems have important differences from the usual theoretical models with purely ferromagnetic exchange and uncorrelated random fields. Specifically, even a minimal model has strong correlations in the random fields between neighbouring pseudospins (via intermediate random charges), while further correlations between those random charges imposed by mesoscopic charge neutrality and competitive effective exchange are anticipated also to be relevant. More generally, alloy-modified $ABO_3$ perovskite systems can involve also other random exchange effects due to ionic replacements. There is need for further study to determine minimal features, but a starting point seems now reasonably clear.

One can certainly conclude that materials science can provide an extremely rich source for many-body systems exhibiting complex macroscopic behaviour in their local displacement correlation behaviour, due to quenched disorder and frustration. One might also note that the temperature range of behaviours of interest in this regard in these structural systems is much higher than those in conventional magnetic systems and the concentration of defects needed to induce strain glass behaviour is much smaller than those for either metallic or semi-conducting spin glasses.

Finally, it is suggested that idealized models conceptually stimulated by these materials systems offer several interesting topics for further fundamental theoretical analysis and computer simulation.

**Acknowledgements**

The author is grateful to Avadh Saxena and Turab Lookman for introducing him to martensitic shape-memory alloys and for useful discussions on the topic over many years of visits to Los Alamos National Laboratory, whose hospitality he also acknowledges. Also, in connection with martensitic alloys, he has appreciated correspondence with Jim Sethna and Xiaobing Ren. He thanks Roger Cowley, his colleague at Oxford, for introducing him to relaxor ferroelectrics, for informing him of several results and comparisons and for useful discussions on how to model and understand, and also Wolfgang Kleemann for very helpful comments on a draft of this paper and for drawing his attention to other relevant works on relaxors. Finally, he apologises again to the experts in martensites and relaxors whose work he has not acknowledged and indeed much of which he is insufficiently familiar with, but if he waited until he had had an opportunity to read, absorb and understand everything that has been done and written about, this article would not have been completed. Hopefully it will stimulate reactions, even if only of correction and objection.

## References

[1] See for example, D.Sherrington: Physics and Complexity, Phil.Trans.Roy.Soc.A **368**, 1175 (2010)
[2] See for example, K.Binder and A.P.Young: Spin glasses: Experimental facts, theoretical concepts, and open questions, Rev.Mod.Phys. **58**, 801 (1986); D. Sherrington: Spin glasses: a Perspective, in eds. E.Bolthausen and A.Bovier, *Spin Glasses* (Springer, Berlin 2006)
[3] See for example, M.Mézard, G.Parisi and M.A.Virasoro: *Spin Glass Theory and Beyond* (World Scientific, Singapore, 1987); H.Nishimori: *Statistical Physics of Spin Glasses and Neural Networks* (Oxford University Press, Oxford, 2001); M.Mézard and A.Montanari: *Information, Physics and Computation* (Oxford University Press, Oxford, 2009)
[4] See for example, M.Talagrand: *Spin glasses: a Challenge for Mathematicians* (Springer, Berlin 2003); A.Bovier: *Statistical Physics of Disordered Systems: a Mathematical Perspective* (Cambridge University Press, Cambridge 2006)
[5] See for example, J.Mydosh: *Spin Glasses: an Experimental Introduction* (Taylor and Francis, London 1995); H.Maletta and W.Zinn, Spin glasses, in eds. K.A.Gschneider Jr. and L.Eyring, *Handbook on the Physics and Chemistry of Rare Earths* **12**, 213 (Elsevier 1989)
[6] See for example, eds.M.Heimel, M.Pleiming and R.Sanctuary: *Ageing and the Glass Transition,* (Springer, Berlin 2007)
[7] B.R.Coles, B.Sarkissian & R.H.Taylor; The role of finite magnetic clusters in Au-Fe alloys near the percolation concentration, Phil.Mag. B **37,** 489 (1978)
[8] H. Maletta & P.Convert; Onset of Ferromagnetism in EuxSr1-xS near x=0.5, Phys.Rev.Lett. 42, 108 (1979)
[9] D.Sherrington and S.Kirkpatrick; Solvable Model of a Spin Glass, Phys.Rev.Lett. **35**, 1972 (1975)
[10] S.Nagata, P.H.Keesom and H.R.Harrison: Low-dc-field susceptibility of **Cu**Mn spin glass, Phys.Rev. B **19**, 1633 (1979)
[11] V.Dupuis, E.Vincent, J.-P.Bouchaud, J.Hammann, A.Ito & H.Aruga Katori: Aging, rejuvenation, and memory effects in Ising and Heisenberg spin glasses, Phys.Rev. B **64**, 174204 (2001)
[12] S.F.Edwards & P.W.Anderson: Theory of Spin Glasses, J.Phys.F **5**, 965 (1975)
[13] See for example K.Bhattacharya, *Microstructure of Martensite* (Oxford University Press, 2003)
[14] D.Sherrington: A Simple Spin Glass Perspective on Martensitic Shape-memory Alloys, J.Phys.Cond.Mat. 20, 304213 (2008)
[15] S.Kartha, T.Castán, J.A.Krumhansl & J.P.Sethna: Spin-glass nature of tweed precursors in martensitic transformations, Phys.Rev.Lett. **67**, 3630 (1991).
[16] S.Kartha, J.A.Krumhansl, J.P.Sethna and L.K.Wickham: Disorder-driven pretransitional tweed pattern in martensitic transformations, Phys.Rev.**B 52**, 803 (1995)
[17] S.R.Shenoy and T.Lookman, Strain pseudospins with power-law interactions: Glassy textures of a cooled coupled-map lattice, Phys.Rev. B**78**, 144103 (2008)
[18] T.Lookman,S.R.Shenoy, K.Ø.Rasmussen, A.Saxena and A.R.Bishop: Ferroelastic dynamics and strain compatibility, Phys.Rev. **B67**, 024114 (2003)
[19] A.Guiliani, J.L.Lebowitz and E.H.Lieb, Ising models with long-range antiferromagnetic and short-range ferromagnetic interactions, Phys. Rev. B 74, 064420 (2006)
[20] M.Porta, T.Castán, P.Lloveras, T.Lookman, A.Saxena & S.R.Shenoy: Interfaces in ferroelastics: Fringing fields, microstructure, and size and shape effects, Phys. Rev. B **79**, 214117 (2009)

[21] See for example, M.R.Garey and D.S. Johnson, *Computers and intractability:a guide to the theory of NP-completeness*, W.H.Freeman (New York 1979)].
[22] S.Sarkar, X.Ren and K.Otsuka, Evidence for Strain Glass in the Ferroelastic-Martensitic System $Ti_{50-x} Ni_{50+x}$, Phys. Rev. Lett. 95, 205702 (2005).
[23] Y.Wang, X.Ren, K.Otsuka and A.Saxena, Evidence for broken ergodicity in strain glass, Phys. Rev. B 76, 132201 (2007)
[24] X.Ren, Y.Wang, K.Otsuka, P.Loveras, T.Castán, M.Porta, A.Planes and A.Saxena, Ferroelastic nanostructures and nanoscale transitions: ferroics with point defects, MRS Bull. **34**, 838 (2009)
[25] N.Schupper & N.M.Scherb, Inverse melting and inverse freezing: a spin model, Phys.Rev. E **72,** 046107 (2005)
[26] N. Gayathri, A.K.Raychaudhuri, S.K.Tiwary, R.Gundakaram, A.Arulraj & C.N.R.Rao: Electrical transport, magnetism, and magnetoresistance in ferromagnetic oxides with mixed exchange interactions: A study of the $La_{0.7}Ca_{0.3}Mn_{1-x}Co_xO_3$ system, Phys. Rev. B **56**, 1345 (1997)
[27] D.Elderfield & D.Sherrington: The curious case of the Potts spin glass, J.Phys. C **16**, 4865 (1983)
[28] X.Ren, Y.Wang, Y.Zhou, Z.Zhang, D.Wang, G.Fan, K. Otsuka, T.Suzuki, Y.Ji, J.Zhang, Y.Tian, S.Hoi, X.Ding: Strain glass in ferroelastic systems: Premartensitic tweed versus strain glass, Phil.Mag. **90**, 141 (2010)
[29] J.P.Sethna and C.R.Myers: Martensitic Tweed and the Two-Way Shape-Memory Effect, arXiv:cond-mat/970203 (1997)
[30] L.E.Cross: Relaxor ferroelectrics, Ferroelectrics **76**, 241 (1987)
[31] W.Kleemann, Random fields in dipole glasses and relaxors, J.Non-Crys.Sol. **307-310**, 66 (2002); W.Kleemann: The relaxor enigma - charge disorder and random fields in ferroelectrics, J. Mater. Science **41**, 129 (2006)
[32] For a recent review see R.A.Cowley, S.N.Gvasaliya, S.G.Lushnikov, B.Roessli & G.M.Rotaru: Relaxing with relaxors: a review of relaxor ferroelectrics; to be published
[33] V.Westphal, W.Kleemann & M.D.Glinchuk: Diffuse phase transitions and random-field-induced domain states of the ''relaxor'' ferroelectric $PbMg_{1/3}Nb_{2/3}O_3$, Phys.Rev.Lett. **68**, 847 (1992)
[34] G.A.Smolenskii, V.A.Isupov, A.I.Agranoyskaya and S,N.Popov: Ferroelectrics with diffuse phase transitions, Sov.Phys.Solid State **2**, 2584 (1961)
[35] G.V.Lecomte, H.von Löhneysen and E.F.Wassermann: Frequency dependent magnetic susceptibilty and spin glass freezing in **Pt**Mn alloys, Z.Phys. B **50**, 239 (1983)
[36] See for example A.P.Young (ed.): *Spin Glasses and Random Fields* (World Scientific, Singapore 1998)
[37] Y.Imry & S-K.Ma: Random-field instability of the ordered state of continuous symmetry, Phys.Rev.Lett. **35**, 13909 (1975)
[38] M.Mėzard and R.Monasson, Glassy transition in the three-dimensional Ising model, Phys.Rev. B **50**, 7199 (1994)
[39] F.Krzakala, F.Ricci-Tersenghi & L.Zdeborová: Elusive Glassy Phase in the Random Field Ising Model, Phys.Rev.Lett.**104**, 207208 (2010)
[40] H.Yoshizawa, R.Cowley, G.Shirane & R.J.Birgenau: Neutron scattering study of the effect of a random field on the three-dimensional dilute Ising antiferromagnet $Fe_{0.6}Zn_{0.4}F_2$, Phys.Rev. B **31,** 4548 (1985); P.Pollak, W.Kleemann and D.P.Belanger: Metastability of the uniform magnetization in three-dimensional random-field Ising model systems. II $Fe_{0.47}Zn_{0.53}F_2$, Phys.Rev. B **38**, 4773 (1988); F.C.Montenegro, A.R.King, V.Jaccarino, S-J.Han and D.P.Belanger: Random-field-induced spin-glass behavior in the diluted Ising antiferromagnet $Fe_{0.31}Zn_{0.69}F_2$, Phys.Rev. B **44**, 2155 (1991)
[41] See for example, S.Bedanta and W.Kleemann, Supermagnetism, J.Phys. D **42**, 013001 (2009); J.F. Fernández, Equilibrium spin-glass transition of magnetic dipoles with random anisotropy axes, Phys.Rev. B **78**, 064404 (2008); J.F.Fernandez and J.J.Alonso: Equilibrium spin-glass transition of magnetic dipoles with random anisotropy axes on a site diluted lattice, Phys.Rev. B **79**, 214424 (2009); and references therein
[42] W.Kleemann, J.Dec, P.Lehnen, R.Blinc, B.Zalar and P.Pankrath: Uniaxial relaxor ferroelectrics:the ferroic random-field Ising model materialized at last, Europhys. Lett. **57**, 14 (2002)
[43] S.K.Ghatak and D.Sherrington: Crystal field effects in a general S Ising spin glass, J.Phys.C **10**, 3149 (1977)
[44] A.Crisanti and L.Leuzzi, Thermodynamic properties of a full-replica-symmetry-breaking Ising spin glass on lattice gas: The random Blume-Emery-Griffiths-Capel model, Phys.Rev. B **70**, 014409 (2004)
[45] R.Vasseur and T.Lookman: Effects of disorder in ferroelastics: a spin glass model for strain glass, Phys.Rev. B **81**, 094107 (2010)

[46] S.Boettcher and A.G.Percus: Optimization with extremal dynamics, Phys.Rev.Lett. **86**, 5211 (2001)

[47] S.Shenoy, T.Lookman, A.Saxena and A.R.Bishop: Martensitic textures: multiscale consequences of elastic compatibility, Phys.Rev.B **60**, R12537 (1999)

[48] See for example [17] and references therein; also M.Vojta; Lattice symmetry breaking in cuprate superconductors: stripes, nematics and superconductivity, Adv.Phys. **58**, 699 (2009)

[49] R.Blinc and R.Pirc: Spherical random-bond–random-field model of relaxor ferroelectrics, Phys.Rev.Lett **83**, 424 (1999).